\documentclass[multphys,vecphys]{svmult}

\usepackage{makeidx}         % allows index generation
\usepackage{graphicx}        % standard LaTeX graphics tool
\usepackage{multicol}        % used for the two-column index
\usepackage[bottom]{footmisc}% places footnotes at page bottom
\makeindex             % used for the subject index
\begin{document}

\title*{Nonquasiparticle states in half-metallic ferromagnets}

\author{V. Yu.~Irkhin$^{1}$, M. I.~Katsnelson$^{2}$, and A. I. Lichtenstein$^{3}$}
\institute{$^{1}$Institute of Metal Physics, 620219, Ekaterinburg, Russia \\
$^{2}$Department of Physics, Uppsala University, Box 530, SE-751 21 Uppsala,%
\\
Sweden\\
$^{3}$Institute of Theoretical Physics, University of Hamburg,
Jungiusstrasse 9, 20355 Hamburg, Germany}
%\texttt{Valentin.Irkhin@imp.uran.ru}
 \maketitle

\begin{abstract}
Anomalous magnetic and electronic properties of the half-metallic
ferromagnets (HMF) have been discussed. The general conception of
the HMF electronic structure which take into account the most
important correlation effects from electron-magnon interactions,
in particular, the spin-polaron effects, is presented. Special
attention is paid to the so called non-quasiparticle (NQP) or
incoherent states which are present in the gap near the Fermi
level and can give considerable contributions to thermodynamic
and transport properties. Prospects of experimental observation of
the NQP states in core-level spectroscopy is discussed. Special
features of transport properties of the HMF which are connected
with the absence of one-magnon spin-flip scattering processes are
investigated. The temperature and magnetic field dependences of
resistivity in various regimes are calculated. It is shown that
the NQP states can give a dominate contribution to the
temperature dependence of the impurity-induced resistivity and in
the tunnel junction conductivity. First principle calculations of
the NQP-states for the prototype half-metallic material NiMnSb
within the local-density approximation plus dynamical mean field
theory (LDA+DMFT) are presented.
\end{abstract}

\section{Introduction}

Half-metallic ferromagnets (HMF) \cite{degroot,IK,pickett}
attract recently a great scientific and industrial attentions due
to their importance for spin-dependent electronics or
``spintronics'' \cite{prinz}. The HMF have metallic electronic
structure for one spin projection (majority- or minority-spin
states), but for the opposite spin direction the Fermi level lies
in the energy gap \cite{degroot}. Therefore the spin-up and
spin-down contributions to electronic transport properties have
different orders of magnitude, which can result in a huge
magnetoresistance for heterostructures containing the HMF
\cite{IK}.

At the same time, the HMF are very interesting conceptually as a
class of materials which may be suitable for investigation of the
essentially many-body physics ``beyond standard band theory''. In
the most cases many-body effects lead only to renormalization of
the quasiparticle parameters in the sense of Landau's Fermi
liquid theory, the electronic {\it liquid} being {\it
qualitatively} similar to the electron {\it gas} (see, e.g.,
\cite{Nozieres,VK}. On the other hand, due to specific band
structure of the HMF, an important role belongs here to incoherent
(nonquasiparticle, NQP) states which occur near the Fermi level
because of correlation effects \cite{IK}. The appearance of NQP
states in the energy gap near the Fermi level is one of the most
interesting correlation effects typical for the HMF. The origin of
these states is connected with ``spin-polaron'' processes: the
spin-down low-energy electron excitations, which are forbidden
for the HMF in standard one-particle scheme, turn out to be
possible as superpositions of spin-up electron excitations and
virtual magnons. The density of these nonquasiparticle states
vanishes at the Fermi level, but increases drastically at the
energy scale of the order of a characteristic magnon frequency
$\overline{\omega }$. The NQP states were first considered
theoretically by Edwards and Hertz \cite{edwards} in the
framework of a broad-band Hubbard model for itinerant electron
ferromagnets. Later it was demonstrated \cite{IK1} that for a
{\it narrow-band} (infinite-$U$) Hubbard model the {\it whole}
spectral weight for one spin projection belongs to the NQP states
which is of crucial importance for the problem of stability of
Nagaoka's ferromagnetism \cite{nagaoka} and for adequate
description of corresponding excitation spectrum. The NQP states
in the $s-d$ exchange model of magnetic semiconductors have been
considered in Ref. \cite{AI}. It was shown that depending on the
sign of the $s-d$ exchange integral, the NQP states can form
either only below the Fermi energy $E_F$ or only above it. Later
it was realized that the HMF are natural substances for
theoretical and experimental investigating of the NQP effects
\cite{IK90}. A variety of these effects in the electronic and
magnetic properties has been considered (for review of the earlier
works see Ref. \cite{IK}) and some recent developments will be
discussed in the present paper. As an example of highly unusual
properties of the NQP states, we note that they can contribute to
the $T$-linear term in the electron heat capacity \cite
{IK90,IKT}, despite their density at $E_F$ vanishes at
temperature $T=0$. Existence of the NQP states at the HMF surface
has been predicted in Ref. \cite{KE} and may be important for
their detection by surface-sensitive methods such as the ARPES
\cite{ARPES} or by spin-polarized scanning tunneling microscopy
\cite{STM}. Recently the density of NQP states has been
calculated from first principles for a prototype HMF, NiMnSb
\cite{lulu}. Some effects of the NQP states on physical
properties of the HMF will be considered below. Because of the
volume restrictions we will concentrate on several examples
skipping the temperature dependence of nuclear magnetic
relaxation rate \cite{NMR} and many others.

\section{Origin of nonquasiparticle states and electron spin polarization in
the gap}

From theoretical point of view, the HMF are characterized by the
absence of magnons decay into the Stoner excitations (pairs
electron-hole with the opposite spins). Therefore spin waves are
well defined in the whole Brillouin zone, similar to the
Heisenberg ferromagnets and degenerate ferromagnetic
semiconductors. Thus, unlike for the usual itinerant ferromagnets,
effects of electron-magnon interactions (so-called spin-polaron
effects) are not masked by the Stoner excitations in the HMF and
may be studied in a ``pure'' form. As we will see below, the
electron-magnon scattering results in the occurrence of NQP
states.

We start our consideration of the interaction of charge carriers
with local moments in the standard $s$-$d$ exchange model
\cite{magnetism}. The $s$-$d$ exchange Hamiltonian reads
\begin{equation}
{\cal H}=\sum_{{\bf k}\sigma }t_{{\bf k}}c_{{\bf k}\sigma }^{\dagger }c_{%
{\bf k}\sigma }-\sum_{{\bf qk}}I_{{\bf k,k+q}}\sum_{\alpha \beta }{\bf S_q}%
c_{{\bf k}\alpha }^{\dagger }\mbox {\boldmath $\sigma $}_{\alpha \beta }c_{%
{\bf k-q}\beta }-\sum_{{\bf q}}J_{{\bf q}}{\bf S}_{{\bf q}}{\bf S}_{-{\bf q}}
\label{H}
\end{equation}
where $c_{{\bf k}\sigma }^{\dagger }$, $c_{{\bf k}\sigma }$ and ${\bf S}_{%
{\bf q}}$ are operators for conduction electrons and localized
spins in the quasimomentum representation, the electron spectrum
$t_{{\bf k}}$ is referred to the Fermi level $E_F$, $I_{{\bf
k,k+q}}$ is the $s$-$d$ exchange parameter, ${\bf \sigma }$ are
the Pauli matrices. We include in the Hamiltonian explicitly the
``direct'' $d$-$d$ exchange interaction (last term in
Eq.(\ref{H})) to construct perturbation theory in a convenient
form. In real materials, this interaction may have a superexchange
nature or result from the indirect exchange via conduction
electrons (in the HMF situation, this is not reduced to the RKKY
interaction). In the latter case, the $d$-$d$ exchange
interaction comes from the same $s$-$d$ interaction and cannot be
considered as an independent parameter. However, as demonstrated
by direct calculations (see e.g. Refs.\cite{AI,AI2}), the
corresponding terms with magnon frequencies occur in higher order
of the $I$ perturbations, for the case where the bare $d$-$d$
exchange interaction is absent.

The $s$-$d$ exchange model does not describe properly the
electronic structure for such HMF as the Heusler alloys or
CrO$_{2}$, where there is no domination of the $sp$-electrons in
electronic transport, and a separation of electrons into a
localized $d$-like and a delocalized $s$-like group is
questionable. In such a case, the Hubbard model which describes
the Coulomb correlations in a $d$-band is more appropriate.
However, qualitative effects of electron-magnon interaction do
not depend on the microscopic model. The calculations of the
electron and magnon Green's functions in the non-degenerate
Hubbard model were performed in Refs. \cite{edwards,IK90} and gave
practically the same result as the $s$-$d$ exchange model with
simple replacement of $I$ by the Hubbard parameter $U$.

%%%%%%%%%%%%%%%%%%%% Fig.1 %%%%%%%%%%%%%%%%%%%%%%%%%%%%%%%%%%%%%%%%%%%%%%%
\begin{figure}
\centering
\includegraphics[height=5cm]{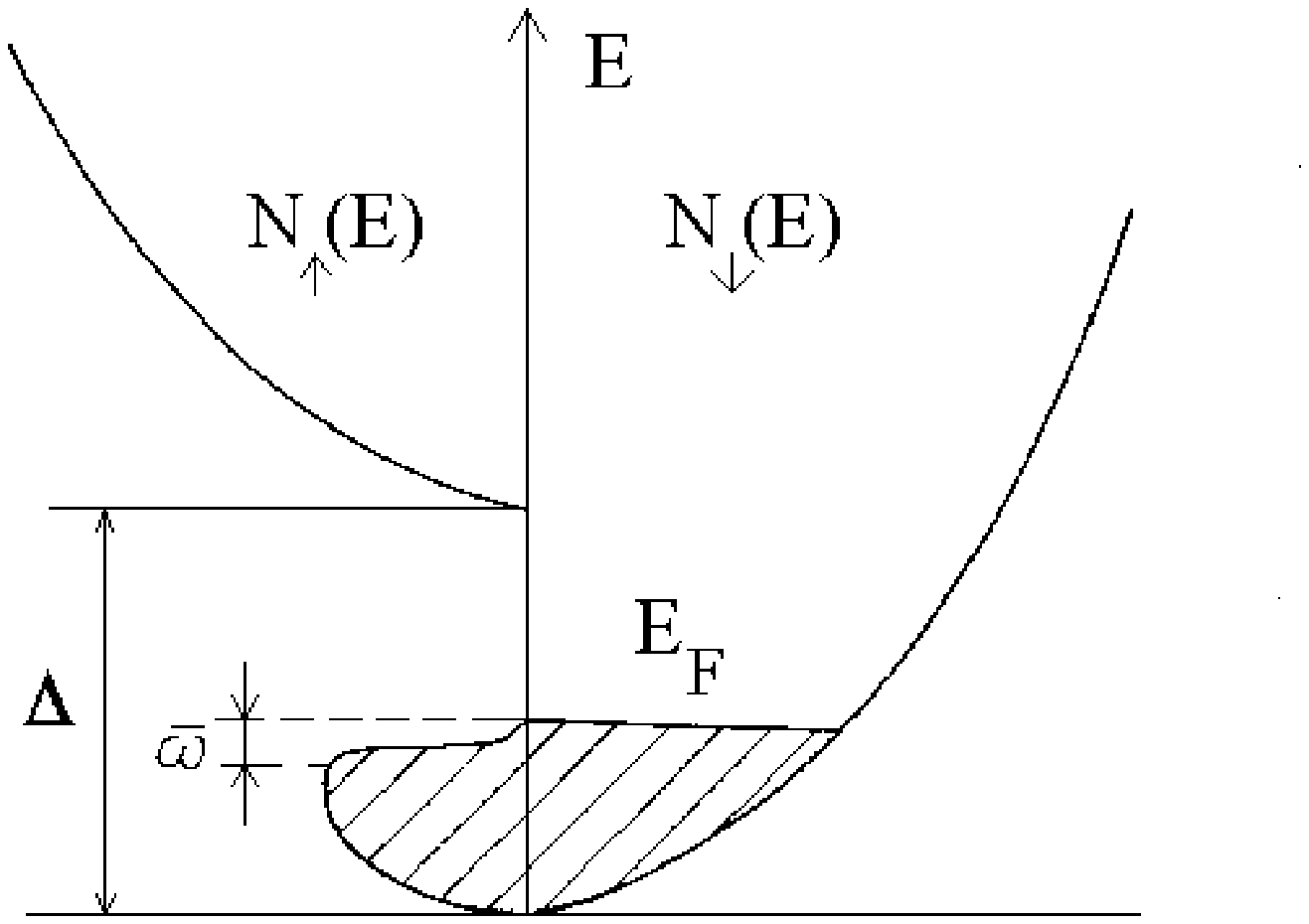}
\caption{Density of states in a half-metallic ferromagnet with
$I<0$ (schematically). Non-quasiparticle states with
$\protect\sigma =\uparrow $ occur below the Fermi level.}
\label{fig:1}
\end{figure}
%%%%%%%%%%%%%%%%%%%%%%%%%%%%%%%%%%%%%%%%%%%%%%%%%%%%%%%%%%%%%%%%%%%%%%%%%%
As demonstrated by analysis of the electron-spin coupling, the
NQP picture turns out to be different for two possible signs of
the $s-d$ exchange parameter $I $. For $I<0$ case, the spin-up NQP
states appears below the Fermi level as an isolated region in the
energy dyagram (Fig. \ref{fig:1}). The occupied states with the
total spin $S-1$ are a superposition of the states $|S\rangle
|\downarrow \rangle $ and $|S-1\rangle |\uparrow \rangle $. The
entanglement of the states of electron and spin subsystems which
is necessary to form the NQP states is a purely quantum effect
formally disappearing at $S\rightarrow \infty $. For qualitative
understanding why the NQP states are formed only below the $E_F$
in this case, we consider a limit $I \rightarrow -\infty$ . Then
the charge carrier is really a many-body state of the occupied
site with total spin $S-1/2,$ which propagates in the
ferromagnetic medium with spin $S$ at any other site. The
fractions of the states $|S\rangle |\downarrow \rangle $ and
$|S-1\rangle|\uparrow \rangle$ in the charge mobile carrier state are
$1/(2S+1)$ and $2S/(2S+1)$, respectively, so that the first
number is just a spectral weight of {\it occupied} spin-up
electron NQP states. At the same time, the density of {\it empty}
states is measured by the number of electrons with a given spin
projection which can added to the system. It is obvious that one
cannot put any spin-up electrons in the spin-up site with
$I=-\infty$. Therefore the density of NQP states should vanish
above the the $E_F$.

\begin{figure}
\centering
\includegraphics[height=5cm]{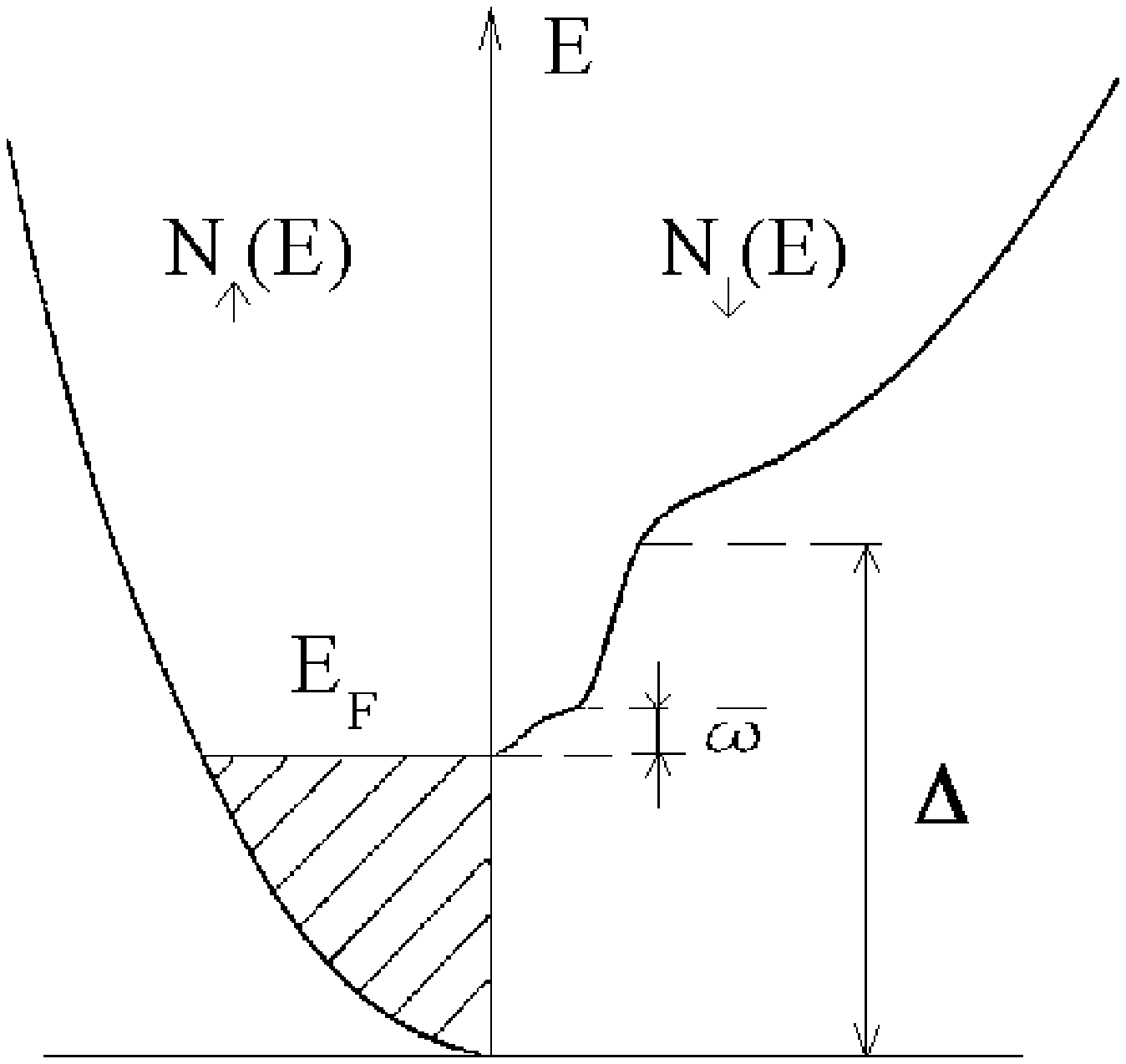}
\caption{Density of states in a half-metallic ferromagnet with
$I>0$ (schematically). Non-quasiparticle states with
$\protect\sigma =\downarrow $ occur above the Fermi level.}
\label{fig:2}
\end{figure}

On contrary, for the $I>0$ case, the spin-down NQP scattering
states form a ``tail'' of the upper spin-down band, which starts
from the $E_F$ (Fig.\ref {fig:2}) since the Pauli principle
prevents electron scattering into occupied states. A similar
analysis of the limit $I\rightarrow +\infty $ helps to understand
the situation qualitatively.
%%%%%%%%%%%%%%%%%%%% Fig.2 %%%%%%%%%%%%%%%%%%%%%%%%%%%%%%%%%%%%%%%%%%%%%%%
%%%%%%%%%%%%%%%%%%%%%%%%%%%%%%%%%%%%%%%%%%%%%%%%%%%%%%%%%%%%%%%%%%%%%%%%%%

It is worthwhile to note that in the most known HMF energy gap
exists for minority-spin states \cite{IK} which is similar to the
case $I>0$, therefore the NQP states should arise {\it above} the
Fermi energy. For exceptional cases with the {\it majority}-spin
gap such as a double perovskite Sr$_{2}$FeMoO$_{6}$ \cite{double}
one should expect the NQP states {\it below} the Fermi energy.
This would be very interesting since in the latter case the NQP
states can be probed by spin-polarized photoemission which is
technically much simpler than spin-polarized BIS spectra
\cite{BIS} needed to probe the empty NQP states.

Let us consider now the density of states (DOS) scheme for the HMF
within the $s-d$ exchange model more quantitatively \cite{IK,AI}.
Neglecting the ${\bf k}$ -dependence of $s-d$ exchange
interaction, the electron Green's function has the following form
\begin{equation}
G_{{\bf k}}^\sigma (E)=\left[ E-t_{{\bf k}\sigma }-\Sigma _{{\bf k}\sigma
}(E)\right] ^{-1}  \label{dys}
\end{equation}
where $t_{{\bf k}\sigma }=t_{{\bf k}}-\sigma I\langle S^z\rangle
$ is the mean-field electron spectrum and $\Sigma _{{\bf %
k}\sigma}(E)$ is the self-energy which describe the
electron-magnon interactions. Within the second order
approximation in $I$ one has  $\Sigma _{{\bf k}\sigma %
}(E)=2I^2SQ_{{\bf k}}^\sigma (E)$ with
\begin{equation}
Q_{{\bf k}}^{\uparrow }(E)=\sum_{{\bf q}}\frac{N_{{\bf q}}+n_{{\bf k+q}%
}^{\downarrow }}{E-t_{{\bf k+q\downarrow }}+\omega _{{\bf q}}} \ , \ \  Q_{{\bf k}%
}^{\downarrow }(E)=\sum_{{\bf q}}\frac{1+N_{{\bf q}}-n_{{\bf k-q}}^{\uparrow
}}{E-t_{{\bf k-q\uparrow }}-\omega _{{\bf q}}}  \label{sigma}
\end{equation}
Below we will present more accurate results for the Green's functions (see
Eq.(\ref{g0})) but here the lowest-order perturbation expression (\ref{sigma}%
) will be sufficient.

Using an expansion of the Dyson equation (\ref{dys}) we obtain a
simple expression for the electron DOS
($ -\frac{1}{\pi }\rm{Im}\sum_{{\bf k}}G_{{\bf k}\sigma }(E)$)
\begin{eqnarray}
N_{\sigma }(E)=\sum_{{\bf k}}\delta (E-t_{{\bf k}\sigma })
-\sum_{{\bf k}}\delta ^{\prime }(E-t_{{\bf k}\sigma })\rm{Re}%
\Sigma _{{\bf k}\sigma }(E)-\frac{1}{\pi }\sum_{{\bf k}}\frac{\rm{Im}%
\Sigma _{{\bf k}\sigma }(E)}{(E-t_{{\bf k}\sigma })^{2}}  \label{N(E)}
\end{eqnarray}
The second term in the right-hand side of Eq. (\ref{N(E)})
describes the renormalization of quasiparticle energies. The
third term, which arises from the branch cut of the self-energy
$\Sigma _{{\bf k}\sigma }(E)$, describes the incoherent
(nonquasiparticle) contribution owing to scattering by magnons.
One can see that the NQP does not vanish in the energy region,
corresponding to the ``alien'' spin subband with the opposite projection $%
-\sigma $. Substituting Eq.(\ref{sigma}) into Eq.(\ref{N(E)}) and
neglecting the quasiparticle shift we obtain for the case of HMF
with $I>0$
\begin{eqnarray}
N_{\uparrow }(E) &=&\sum_{{\bf kq}}\left[ 1-\frac{2I^{2}SN_{{\bf q}}}{(t_{%
{\bf k+q\downarrow }}-t_{{\bf k\uparrow }})^{2}}\right] \delta (E-t_{{\bf k}%
\uparrow })  \nonumber \\
N_{\downarrow }(E) &=&2I^{2}S\sum_{{\bf kq}}\frac{1+N_{{\bf q}}-n_{{\bf %
k\uparrow }}}{(t_{{\bf k+q\downarrow }}-t_{{\bf k\uparrow }}-\omega _{{\bf q}%
})^{2}}\delta (E-t_{{\bf k}\uparrow }-\omega _{{\bf q}})  \label{DOS1}
\end{eqnarray}

\begin{figure}
\centering
\includegraphics[height=5cm]{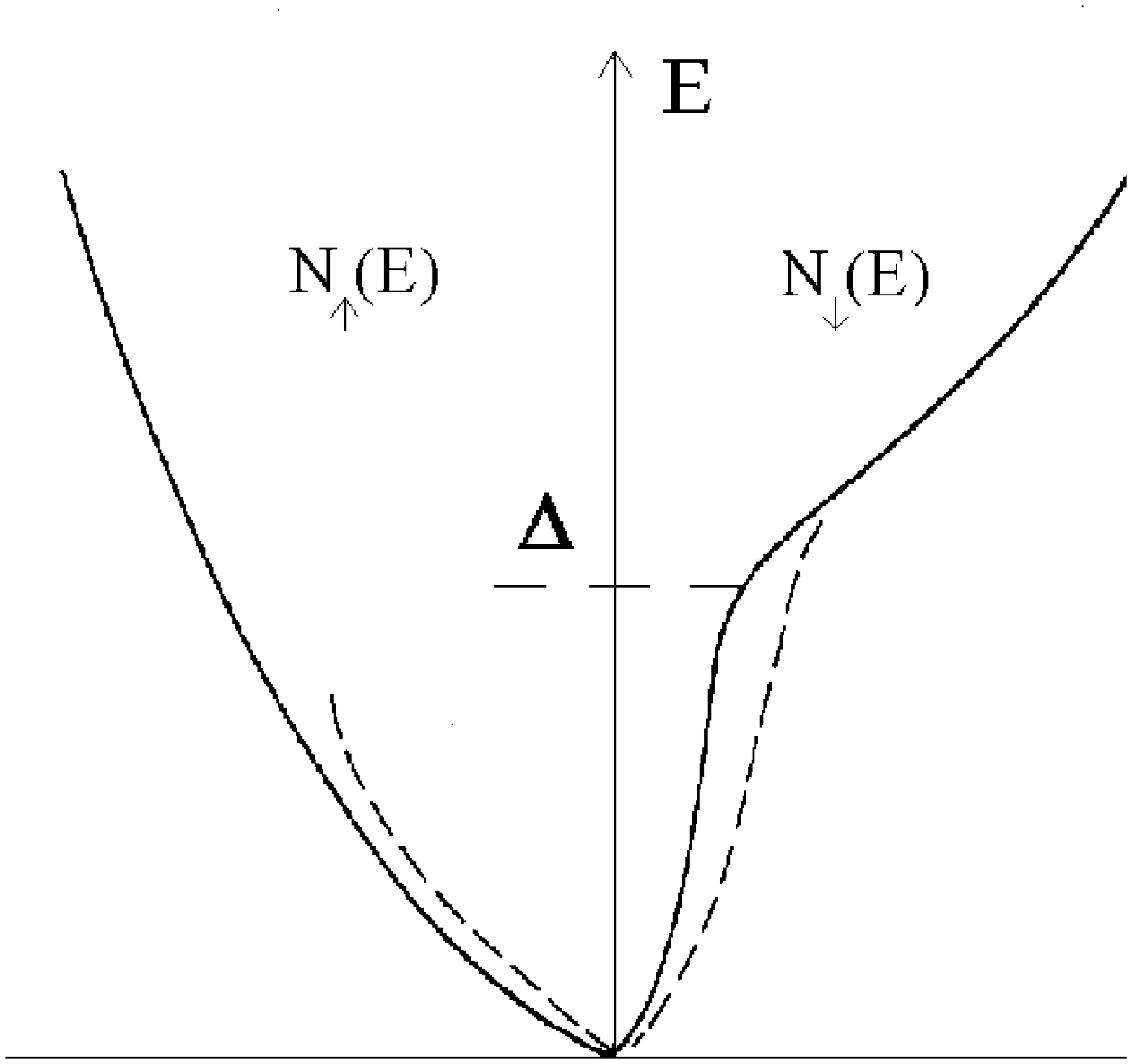}
\caption{Density of states in the s-d model in the case of empty
conduction band ($I > 0$). At $T = 0$ (solid line) the
spin-polaron tail of spin-down states reaches the band bottom.
The dashed line corresponds to finite temperatures.} \label{fig:3}
\end{figure}

The DOS for case of the empty conduction band is shown in
Fig.\ref{fig:3}. The $T^{3/2}$-dependence of the magnon
contribution to the residue of the Green's function (\ref{dys}),
which follows from (\ref{sigma}), i.e. of the effective electron
mass in the lower spin subband, and an increase with temperature
of the incoherent tail from the upper spin subband result in a
strong temperature dependence of partial densities of states
$N_\sigma (E)$, the corrections being of opposite sign.
%%%%%%%%%%%%%%%%%%%% Fig.3 %%%%%%%%%%%%%%%%%%%%%%%%%%%%%%%%%%%%%%%%%%%%%%%
%%%%%%%%%%%%%%%%%%%%%%%%%%%%%%%%%%%%%%%%%%%%%%%%%%%%%%%%%%%%%%%%%%%%%%%%%%

The behaviour of $N(E)$ near the Fermi level in the HMF (or
degenerate ferromagnetic semiconductor) turns out to be also
non-trivial (Figs. 2,3). If we neglect magnon frequencies in the
denominators of Eq.(\ref{DOS1}), the partial density of
incoherent states should occur as a jump above or below the Fermi
energy $E_F$ for the case of $I>0$ and $I<0$ respectively owing
to the Fermi distribution functions. An account of finite magnon
frequencies $\omega _{{\bf q}}=Dq^2$ ($D$ is the spin stiffness
constant) leads to the smearing of these singularities on the
energy interval $\overline{\omega }\ll E_F$, with the $N(E_F)$
being equal to zero. For $|E-E_F|\ll \overline{\omega }$ we obtain
\begin{equation}
\frac{N_{-\alpha }(E)}{N_\alpha (E)}=\frac 1{2S}\left| \frac{E-E_F}{%
\overline{\omega }}\right| ^{3/2}\theta (\alpha (E-E_F)),
\label{alpha1}
\end{equation}
where $\alpha =\rm{sign}(I)=\pm 1$ is the spin projections
$\uparrow ,\downarrow$ of corresponding NQP-states. With increasing $|E-E_F|$,
$N_{-\alpha}/N_\alpha$ tends to a constant value which is of
order of $I^2$ within the perturbation theory. In the strong
coupling limit where $|I|\rightarrow \infty $ we have for the
$|E-E_F|\gg \overline{\omega }$
\begin{equation}
\frac{N_{-\alpha}(E)}{N_\alpha(E)}=\frac1{2S}\theta(\alpha(E-E_F))
\label{alpha2}
\end{equation}

In a simple $s$-$d$ model case, qualitative considerations
\cite{edw}, as well the Green's functions calculations
 \cite{AI,aus1}, gives a spin polarization
of conduction electrons in the spin-wave region proportional
to magnetization:
\begin{equation}
P\equiv \frac{N_{\uparrow }-N_{\downarrow }}{N_{\uparrow }+N_{\downarrow }}%
=2P_{0}\langle S^{z}\rangle   \label{polar1}
\end{equation}
A weak ground-state depolarization $1-P_{0}$ occurs in the case of
$I>0$. The behavior $P(T)\simeq \langle S^{z}\rangle $ is
qualitatively confirmed by experimental data on field emission for
ferromagnetic semiconductors \cite{kisker} and transport
properties for the half-metallic Heusler alloys \cite {otto}.
Note, that the Eq. \ref{polar1} is valid for a hole spin-wave region
only for the narrow-band case (large $I$), whereas for the case of a small $I$
it described spin-polarization only for very low temperatures.

An attempt to generalize the result (\ref{polar1}) to the HMF
case have been made on the basis of qualitative arguments for the
atomic limit \cite{skomski}. We will demonstrate that the
situation for the HMF is more complicated. Let us focus on the
magnon contribution to the DOS (\ref {DOS1}) and calculate a
following function:
\begin{equation}
\Phi =\sum_{{\bf kq}}\frac{2I^2SN_{{\bf q}}}{(t_{{\bf k+q\downarrow }}-t_{%
{\bf k\uparrow }}-\omega _{{\bf q}})^2}\delta (E_F-t_{{\bf k}\uparrow })
\label{phi}
\end{equation}
Using the parabolic electron spectrum $t_{{\bf k\uparrow
}}=k^2/2m$ and averaging over angles of the vector ${\bf k}$, we
obtain
\begin{equation}
\Phi =\frac{2I^2Sm^2}{k_F^2}\rho \sum_{{\bf q}}\frac{N_{{\bf
q}}}{\left( q^{*}\right) ^2-q^2} , \label{phi1}
\end{equation}
where $\rho =N_{\uparrow }(E_F,T=0)$. We have used the condition
$q\ll k_F$, $q^{*}=m\Delta /k_F=\Delta /v_F$, where $\Delta
=2\left| I\right| S$ is the spin splitting$.$ Corresponding
crossover energy scale is equal to $T^{*}=D\left(%
q^{*}\right)^2\sim (\Delta /v_F)^2T_C.$ Finally, we have the
following expression for $\Phi$
\begin{equation}
\Phi =\frac{I^2S}{k_F^2}\frac{m^2}{2\pi ^2}\rho \int_0^\infty \frac{x^{1/2}dx%
}{\exp x-1}\frac 1{T^{*}/T-x}  \label{phi2}
\end{equation}
At the very low temperatures $T<T^{*}$, this result is in
agreement with the qualitative considerations presented above:
\begin{equation}
\Phi =\frac{S-\langle S^z\rangle }{2S}\rho \propto \left( \frac T{T_C}%
\right) ^{3/2}\rho  \label{phi3}
\end{equation}
Nevertheless, for $T>T^{*}$ we have absolutely different
temperature dependence of the spin polarization:
\begin{equation}
\Phi =1.29\frac{\left( q^{*}\right) ^3}{4S\pi ^2}\left( \frac T{T^{*}}%
\right) ^{1/2}\rho  \label{phi4}
\end{equation}

This conclusion is rather important since the crossover
temperature $T^{*}$ can be small and a simple estimation
(\ref{polar1}) may be valid only for very low temperatures.
Moreover, it turns out that the temperature dependence of the
polarization at $T>T^{*}$ is not universal at all. Note that the
model of rigid spin splitting used above is not applicable for
real HMF where the gap has hybridization origin
\cite{degroot,IK}, in contrast to the case of degenerate
ferromagnetic semiconductors. The simplest model for the HMF
consists of a ``normal'' metallic spectrum for the majority
electrons and a hybridization gap for the minority ones
($ \xi _{{\bf k}}  \equiv   t_{{\bf k}\uparrow }-E_F $)
\begin{eqnarray}
t_{{\bf k}\uparrow }-E_F =\frac{k^2-k_F^2}{2m}
\ , \ \
t_{{\bf k}\downarrow }-E_F=\frac 12\left( \xi _{{\bf k}}+{\rm
sgn}\left( \xi _{{\bf k}}\right) \sqrt{\xi _{{\bf k}}^2+\Delta
^2}\right) , \label{model}
\end{eqnarray}
where we assume for simplicity that the Fermi energy lies exactly in the
middle of the hybridization gap. Otherwise one needs to shift $\xi _{{\bf k}%
}\rightarrow \xi _{{\bf k}}+E_0-E_F$ in the last equation, $E_0$
being the middle of the gap. Further, in the expression for
$t_{{\bf k+q\downarrow }}$ one can replace $\xi _{{\bf k+q}}$ by
${\bf v}_{{\bf k}}{\bf q}$, ${\bf v}_{{\bf k}}={\bf k}/m,$ and
use the fact that $\xi _{{\bf k}}=0$ owing to the delta-function
in the definition of $\Phi $. Since a small $q$ give the main
contribution to the estimated integral, we can assume $I_{{\bf %
k,k+q}}\simeq I_{{\bf k,k}}.$ Then one has the following
expression
\begin{eqnarray}
\Phi =2S\sum_{{\bf kq}}I_{{\bf k,k}}^2N_{{\bf q}}\delta (\xi _{{\bf k}%
})\Lambda _{{\bf kq}} \ , \ \ \Lambda _{{\bf kq}} =-\frac
\partial {\partial \omega _{{\bf q}}}\left.
\left\langle \frac 1{t_{{\bf k+q\downarrow }}-t_{{\bf k\uparrow }}-\omega _{%
{\bf q}}}\right\rangle \right| _{\omega _{{\bf q}}=0}
\label{phi10}
\end{eqnarray}
where the angular brackets means the avere over angles of the
vector ${\bf k} $. Simple calculations gives the final result:
\begin{equation}
\Lambda _{{\bf kq}}=\frac 8{v_Fq\Delta }\left( \frac 23\left[
X^3-(X^2+1)^{3/2}+1\right] +X\right),  \label{lambda}
\end{equation}
where $X=k_Fq/m\Delta \equiv q/q^{*}$ ($q^{*}$ is {\it linear} in $\Delta $).
At $X\gg 1$ corresponding to $T\gg T^{*}=Dq^{*2}$, one has, instead of Eq.(\ref{phi4}%
), the following estimation
\begin{equation}
\Phi =\sum_{{\bf kq}}2I^2SN_{{\bf q}}\delta (\xi _{{\bf k}})\frac{16}{%
3v_Fq\Delta }\propto q^{*}\sum_{{\bf q}}\frac{N_{{\bf q}}}q\propto \frac{%
T^{*1/2}}{T_C^{1/2}}T\ln \frac T{T^{*}}  \label{phi11}
\end{equation}
At $X\ll 1$ ($T\ll T^{*}$) we get an universal $T^{3/2}$ behavior
\begin{equation}
\Phi =\rho \sum_{{\bf q}}N_{{\bf q}}\propto \frac{T^{3/2}}{T_C^{1/2}}
\label{phi12}
\end{equation}

The density of NQP states is zero at the Fermi energy only for
$T=0$, while for finite temperatures it is proportional to the
following integral
\begin{equation}
N(E_{F})\propto \int_{0}^{\infty }d\omega \frac{\overline{K}(\omega )}{%
\sinh (\omega /T)} ,  \label{atFermi}
\end{equation}
where $\overline{K}(\omega )$ is a spectral density of the spin
fluctuations \cite{edw,AI,IKT}. Generally speaking, for
temperatures which are comparable with the Curie temperature
$T_{C}$ there are no essential difference between the
half-metallic and ``ordinary'' ferromagnets since a gap of the HMF
is filled. Corresponding analysis for a model of conduction
electrons interacting with ``pseudospin'' excitations in
``ferroelectric'' semiconductors is performed in Ref. \cite{IKT}.
Symmetrical part of the $N(E)$ with respect to the $E_{F}$ in the
gap can be attributed to the smearing of electronic states by the
electron-magnon scattering, while asymmetrical one is the density
of NQP states due to the Fermi distribution function. Note that
this filling of the gap is very important for possible
applications of the HMF in spintronics: they really have some
advantages only in the region of $T\ll T_{C}$. Since a
single-particle Stoner-like theory leads to much less restrictive,
but unfortunately completely wrong condition $T\ll \Delta $, a
many-body treatment of the HMF problem is inevitable.

\section{First-principle calculations of nonquasiparticle states: a
dynamical mean field theory}

A history of the HMF starts from the band-structure of semi
Heusler alloy NiMnSb \cite{degroot}. Later numerous
first-principle electronic structure investigations of the HMF
have been carried out (see, e.g., recent papers
\cite{wijs,weht,freeman,deder} and a review of early works in Ref.
\cite{IK}). All of them are based on a standard local density
approximation (LDA) or generalized gradient approximation (GGA) to
the density functional theory, and, sometimes, on the LDA+U
approximation (see Ref. \cite {korotin} for CrO$_2$). Of course,
essential correlation effects such as NQP states cannot be
considered in these techniques.

Recently, a successful approach has been proposed
\cite{AnisDMFT,lda++} to include correlation effects into the
first-principle electronic structure calculations by combining
the LDA scheme with the dynamical mean-field theory (DMFT). The
DMFT maps a lattice many-body system onto quantum impurity models
subject to a self-consistent condition (for a review, see
Ref.\cite{GKKR}). In this way, the complex lattice many-body
problem splits into simple one-body crystal problem with a local
self-energy and the effective many-body impurity problem. In a
sense, the approach is complementary to the local density
approximation \cite{hohenberg64,kohn65,hedin} where the many-body
problem splits into one-body problem for a crystal and many-body
problem for {\it homogeneous} electron gas. Naively speaking, the
LDA+DMFT method \cite{AnisDMFT,lda++} treats localized $d$- and
$f$-electrons in spirit of the DMFT and delocalized
$s,p$-electrons in spirit of the LDA. Due to numerical and
analytical techniques developed for solution  the effective
quantum-impurity problem \cite{GKKR}, the DMFT become  a very
efficient and extensively used approximation for local energy
dependent self-energy $\Sigma (\omega )$. The accurate LDA+DMFT
scheme can be used for calculating a large number of systems with
different strength of electron correlations (for detailed
description of the method and computational results, see Refs.
\cite{lichtenstein01,held02,kotliar01}). Following the recent
work \cite{lulu} we present here first LDA+DMFT results for the
electronic structure calculations of a ``prototype''
half-metallic ferromagnet NiMnSb.

Before considering the real HMF case, it is worthwhile to check
the applicability of DMFT scheme for quantitative description of
the NQP states. The DMFT is considered as an {\it optimal} local
approximation which means that the self-energy depends only on the
energy and not on the quasimomentum \cite{GKKR}. At the same
time, the NQP states are connected with the self-energy
(\ref{sigma}) which is almost local. It will be {\it exactly}
local if we neglect magnon energies in comparison with the
electron bandwidth, which is rather accurate approximation for
realistic parameters. The local approximation means formally that
we replace the ${\bf q}$-dependent magnon spectral density by the
average one, as in the Eq.(\ref {atFermi}). Such a procedure has
been analyzed and justified in the Ref. \cite{Zarub}. It should be
stressed that an accurate description of the {\it magnon} spectrum
is not important for existence of the NQP states as well as for
proper estimation of their spectral weight, but can be important
for an explicit shape of the DOS tail in the vicinity of the Fermi
level (see Eq.(\ref{alpha1})).

Let us start from the DMFT calculations for the one-band Hubbard
Hamiltonian
\begin{equation}
H=-\sum_{i,j,\sigma }t_{ij}(c_{i\sigma }^{\dagger }c_{j\sigma
}+c_{j\sigma }^{\dagger }c_{i\sigma })+U\sum_in_{i\uparrow
}n_{i\downarrow }, \label{Hubbard}
\end{equation}
on the Bethe lattice with coordination $z\rightarrow \infty $ and
nearest-neighbor hoping $t_{ij}=t/\sqrt{z}$ (in this limit the
DMFT is formally exact \cite{GKKR}). In this case the DOS have a
semicircular form:
\begin{equation}
N(\epsilon )=\frac 1{2\pi t^2}\sqrt{4t^2-\epsilon ^2}  \label{semiel}
\end{equation}
In order to stabilize the HMF state in our toy model, we have
added an external magnetic spin splitting $\Delta$, which mimics
the local Hund polarization from other electrons in the real
NiMnSb compound. This HMF state corresponds to a mean-filed
(Hartree-Fock) solution with a LSDA-like DOS (Fig. (\ref{model}).
%%%%%%%%%%%%%%%%%%%% Fig.6 %%%%%%%%%%%%%%%%%%%%%%%%%%%%%%%%%%%%%%%%%%%%%%%
\begin{figure}
\centering
\includegraphics[height=5cm]{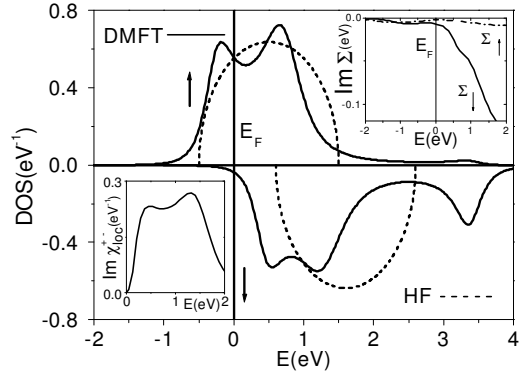}
\caption{ Density of states for HMF in the Hartree-Fock (HF)
approximation (dashed line) and the QMC solution of DMFT problem
for semi-circular model (solid line) with the band-width $W=2$
eV, Coulomb interaction $U=2$ eV, spin-splitting $\Delta=0.5$ eV,
chemical potential $\mu=-1.5$ eV and temperature $T=0.25$ eV.
Insets: imaginary part of the local spin-flip susceptibility
(left) and the spin-rezolved selfenergy (right).} \label{model}
\end{figure}
%%%%%%%%%%%%%%%%%%%%%%%%%%%%%%%%%%%%%%%%%%%%%%%%%%%%%%%%%%%%%%%%%%%%%%%%%%

We can study an average magnon spectrum in this model through the
two-particle correlation function. The local spin-flip
susceptibility
\begin{equation}
\chi _{loc}^{+-}(\tau )=\langle S^{+}(\tau )S^{-}(0)\rangle
=\langle c_{\uparrow }^{\dagger }(\tau )c_{\downarrow }(\tau
)c_{\downarrow }^{\dagger }(0)c_{\uparrow }(0)\rangle ,
\end{equation}
represents the response function required. We have calculated
this function using the numerically exact QMC procedure
\cite{Jarrell92}.

The model DMFT results are presented in Fig. \ref{model}. In
comparison with a simple Hartree-Fock solution (dashed line) one
can see an additional well-pronounced states appearing in the
spin-down gap region, just above the Fermi level. This new
many-body feature corresponds to the NQP states. In addition to
these states visible in both spin channels of the DOS around 0.5
eV, a many-body satellite appears at the energy of 3.5 eV.

The left inset in the Fig. \ref{model} represents the imaginary
part of local spin-flip susceptibility. One can see a well
pronounced shoulder (around 0.5 eV), which is connected with an
average magnon DOS. In addition there is a broad maximum (at 1eV)
corresponding to the Stoner excitation energy. The right inset in
the Fig. \ref{model} represents the imaginary part of self-energy
calculated from our ``toy model''. The spin up channel can be
described by a Fermi-liquid type behavior with a parabolic energy
dependence $-{\rm Im}\Sigma ^{\uparrow }\propto (E-E_F)^2$,
whereas in the spin down channel the imaginary part $-{\rm %
Im}\Sigma ^{\downarrow }$ shows the 0.5eV nonquasiparticle
shoulder. Due to the relatively high temperature of our QMC
calculation (an exact enumeration, technique with the number of
time-slices equal to $L=24$) the NQP tail goes a bit below the
Fermi level, in agreement with Eq.(\ref{atFermi}); at temperature
$T=0$ the NQP tail should ends exactly at the Fermi level.

Let us move to the calculations for real material - NiMnSb. The
details of computational scheme have been described in the Ref.
\cite{lulu}, and only the key points will be mentioned here. In
order to integrate the DMFT approach into the band structure
calculation the so called exact muffin-tin orbital method (EMTO)
\cite {Andersen,EMTO} was used. In the EMTO approach, the
effective one-electron potential is represented by the optimized
overlapping muffin-tin potential, which is the best possible
spherical approximation to the full one-electron potential. The
implementation of the DMFT scheme in the EMTO method is described
in detail in the Ref. \cite{EMTODMFT}. One should note that in
addition to the usual self-consistency of the many-body problem
(self-consistency of the self-energy), a charge self-consistency
has been achieved \cite{kotliar01}.

For the interaction Hamiltonian, a most general rotationally
invariant form of the generalized Hubbard Hamiltonian has been
used \cite{lda++}. The effective many-body impurity problem is
solved using the spin polarized $T$-matrix plus
fluctuation-exchange appriximation (a so-called SPTF)scheme
proposed in the Ref. \cite{Katsnelson02}, which is a development
of the earlier approach \cite{lda++}. The SPTF approximation is a
multiband spin-polarized generalization of a well-known
fluctuation exchange (FLEX) approximation \cite{bickers89}, but
with a different treatment of the particle-hole (PH) and
particle-particle (PP) channels. The particle-particle (PP)
channel is described by a $T$-matrix approach \cite{galitski63}
yielding a renormalization of the effective interaction. The
static part of this effective interaction is used explicitly in
the particle-hole channel.

%%%%%%%%%%%%%%%%%%%% Fig.7 %%%%%%%%%%%%%%%%%%%%%%%%%%%%%%%%%%%%%%%%%%%%%%%
\begin{figure}
\centering
\includegraphics[height=7cm]{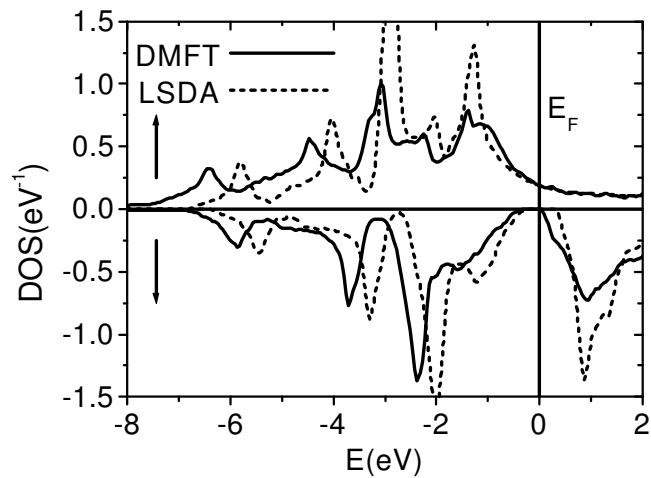}
\caption{Density of states for HMF NiMnSb in LSDA scheme (dashed
line) and in LDA+DMFT scheme (solid line) with  effective Coulomb
interaction $U$=3 eV, exchange parameter $J$=0.9 eV and
temperature $T$=300 K. The nonquasiparticle state is evidenced
just above the Fermi level.} \label{nimnsb}
\end{figure}
%%%%%%%%%%%%%%%%%%%%%%%%%%%%%%%%%%%%%%%%%%%%%%%%%%%%%%%%%%%%%%%%%%%%%%%%%%
There are various methods to estimate the required values of the
on-site Coulomb repulsion energy $U$ and exchange interaction
energy $J$ for realistic materials. The constrained LDA
calculation \cite {AnisimovU} estimates an average Coulomb
interaction between the Mn $d$ electrons as $U=4.8$ eV with an
exchange interaction energy of $J=0.9$ eV. However, this method is
adequate for a typical insulating screening and in general is not
accurate for a metallic kind of screening. The latter will lead
to a smaller value of $U$. Unfortunately, there are no reliable
schemes to calculate $U$ for metals, therefore the results for
different values of $U$ in the energy interval from $0.5$ eV to
the constrained LDA value $U=4.8$ eV have been tested. At the same
time, the results of constrained LDA calculations for the Hund
exchange parameter $J$ do not depends on metallic screening and
should be reliable enough. It appeared that the LDA+DMFT results
are not very sensitive to the value of $U$, due to the $T$-matrix
renormalization. Fig. \ref{nimnsb} represents the results for DOS
using LSDA and LDA+DMFT (with $U=3$ eV and $J=0.9$ eV) approaches.

It is important to mention that the magnetic moment per formula
unit is not sensitive to the $U$ values and is equal exactly $\mu
=4$ $\mu _{B}$, which suggests that the half-metallic state is
stable with respect to the introduction of the correlation
effects. In addition, the DMFT gap in the spin down channel,
defined as the distance between the occupied part and the starting
point of nonquasiparticle state's ``tail'', is also not very
sensitive to the $U$ values. For different $U$'s a slope of the
``tail'' is slightly changed, but the total DOS is weakly
$U$-dependent due to the same $T$-matrix renormalization effects.

%%%%%%%%%%%%%%%%%%%% Fig.8 %%%%%%%%%%%%%%%%%%%%%%%%%%%%%%%%%%%%%%%%%%%%%%%
\begin{figure}
\centering
\includegraphics[height=5cm]{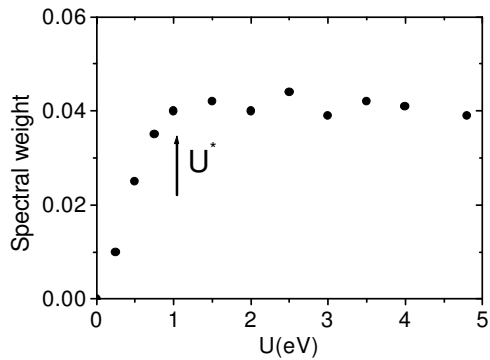}
\caption{Spectral weight of the nonquasiparticle state,
calculated as function of average on-site Coulomb repulsion $U$
at temperature $T$=300 K.} \label{spec}
\end{figure}
%%%%%%%%%%%%%%%%%%%%%%%%%%%%%%%%%%%%%%%%%%%%%%%%%%%%%%%%%%%%%%%%%%%%%%%%%%
Thus the correlation effects do not effect too strongly a general
feature of the electron energy spectrum (except for smearing of
DOS which is due to the finite temperature $T=300$ K in our
calculations). The only qualitatively new effect is the
appearance of a ``tail'' of the NQP states in the energy gap above
the Fermi energy. Their spectral weight for realistic values of
the parameters is not very small, which means that the NQP should
be well pronounced in the corresponding experimental data. A
relatively weak dependence of the NQP spectral weight on the $U$
value (Fig. \ref{spec}) is also a consequence of the $T$-matrix
renormalization \cite{Katsnelson02}. One can see that the
$T$-matrix depends slightly on $U$ provided that the latter is
larger than the widths of the main DOS peaks near the Fermi level
in an energy range of $2$ eV (this is of the order of
$U^{*}\simeq 1$ eV).

\begin{figure}
\centering
\includegraphics[height=5cm]{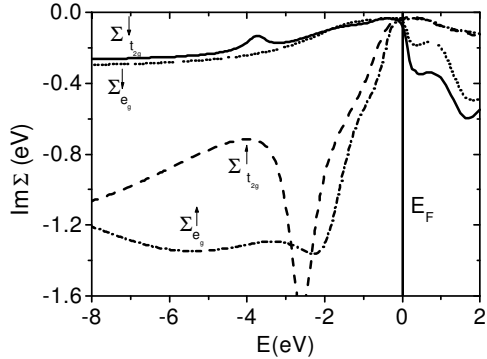}
\caption{The imaginary part of self-energies $
Im\Sigma_d^{\downarrow}$ for $ t_{2g}$ (solid line) and $e_g$
(dotted line), $ Im\Sigma_d^{\uparrow}$ for $t_{2g}$ (dashed
line) and $e_g$ (dashed dotted line) respectively.} \label{self}
\end{figure}

For the spin-up states we have a normal Fermi-liquid behavior
$-{\rm Im}\Sigma _d^{\uparrow }(E)\propto (E-E_F)^2$ with a
typical energy scale of the order of several eV. The spin-down
self-energy behaves in a similar way below the Fermi energy, with
a slightly smaller energy scale (which is still larger than 1
eV). At the same time, a significant increase in ${\rm Im}\Sigma
_d^{\downarrow }(E)$ with a much smaller energy scale (few tenths
of eV) occurs just above the Fermi level, which is more pronounced
for $t_{2g}$ states (Fig. \ref{self}). The similar behavior of
the imaginary part of electronic self-energy and the DOS just
above Fermi level is a signature of the NQP states and is also
noticed in the model calculation (Fig. \ref{model}).
%%%%%%%%%%%%%%%%%%%% Fig.9 %%%%%%%%%%%%%%%%%%%%%%%%%%%%%%%%%%%%%%%%%%%%%%%
%%%%%%%%%%%%%%%%%%%%%%%%%%%%%%%%%%%%%%%%%%%%%%%%%%%%%%%%%%%%%%%%%%%%%%%%%%

Thus the main results of the Ref. \cite{lulu} are ({\it i}) the
existence of the NQP states in real electronic structure of a
specific compound, and ({\it ii}) estimation of their spectral
weight in the LDA+DMFT approach. The temperature dependence of
the NQP density of states in the gap, which is important for
possible applications of the HMF in spintronics, can be analyzed
in the present technique.

\section{X-ray absorption and emission spectra. Resonant x-ray scattering}

Now we discuss the manifestations of NQP states in the core level
spectroscopy \cite{xrayour}. Various spectroscopy techniques such
as x-ray absorption, x-ray emission, and photoelectron
spectroscopies (xas, xes, and xps) give an important information
about the electronic structure of the HMF and related compounds,
such as ferromagnetic semiconductors and colossal
magnetoresistance materials (see, e.g., Refs.
\cite{yarm,yab,kurm,ola}). It is well known that the many-body
effects, particularly dynamical core hole screening, may be
important for a core level spectroscopy even in the case that a
system is not strongly correlated in the initial state
\cite{Mah,noz}. Therefore it is very interesting to look on the
interplay of these effects with the NQP states, which are of
essentially many-body origin themselves.

To consider a core level problem in the HMF we use the same
Hamiltonian of the $s-d$ exchange model, Eq. (\ref{H}) in the
presence of the external potential $U$ induced by the core hole:
\begin{equation}
{\cal H}^{\prime }=\varepsilon _{0}f^{\dagger }f-U\sum_{{\bf
kk}^{\prime }\sigma }c_{{\bf k}\sigma }^{\dagger }c_{{\bf
k}^{\prime }\sigma }f^{\dagger }f , \label{H'}
\end{equation}
where $f^{\dagger },f$ are core hole operators. It is useful to
write down the equation of motion for the retarded two-particle
Green's function \cite{zubarev}
\begin{equation}
G_{{\bf kk}^{\prime }}^{\sigma }(E)=\langle \langle c_{{\bf
k}\sigma }f|f^{\dagger }c_{{\bf k}^{\prime }\sigma }^{\dagger
}\rangle \rangle _{E},
\end{equation}
which determines x-ray absorption and emission spectra
\cite{Mah}. Using a magnon representation for the spin operators,
we derive the following equation for two-particle Green's
function:
\begin{equation}
(E-t_{{\bf k}\sigma })G_{{\bf kk}^{\prime }}^{\sigma }(E)=(1-n_{f}-n_{{\bf k}%
}^{\sigma })\left[ \delta _{{\bf kk}^{\prime }}-U\sum_{{\bf p}}G_{{\bf pk}%
^{\prime }}^{\sigma }(E)\right] -I\sum_{{\bf r}}F_{{\bf k-r,r,k}^{\prime
}}^{\sigma }(E)  \label{g}
\end{equation}
where $n_{f}$ is the occupation number for the $f$-hole in the
initial state, which is further on will be put to zero and $E$ is
the electron energy with respect to $\varepsilon _{0}$). We will
take into account the occupation numbers $n_{{\bf k}}^{\sigma }$
in a simple ladder approximation which works well in the limit of
small concentrations of mobile carriers, except for the immediate
vicinity of the Fermi edge. Here, we do not treat the problem of
the x-ray edge singularity where more advanced approaches are
necessary \cite{Mah,noz}. The following notation has been used in
Eq.(\ref{g}):
\begin{equation}
F_{{\bf k-p,q,k}^{\prime }}^\sigma (E)=(2S)^{1/2}\langle \langle b_{{\bf q}%
}^\sigma c_{{\bf k}-{\bf p,}-\sigma }f|f^{\dagger }c_{{\bf k}^{\prime
}\sigma }^{\dagger }\rangle \rangle _E,
\end{equation}
where $b_{{\bf q}}^{+}=b_{-{\bf q}}^{\dagger }$, $b_{{\bf %
q}}^{-}=b_{{\bf q}}$ are the Holstein-Primakoff magnon operators \cite{magnetism}.
The Green's function $F$ satisfies the equation
\begin{eqnarray}
(E-t_{{\bf k-p,}-\sigma }+\sigma \omega _{{\bf q}})F_{{\bf k-p,q,k}%
^{\prime }}^\sigma (E)
=-U(1-n_{{\bf k-p}}^{-\sigma })\Psi _{{\bf q,k}^{\prime }}^\sigma (E)  \nonumber
\\
-I(N_{{\bf q}}^\sigma +\sigma n_{{\bf k-p}}^{-\sigma })[2SG_{{\bf
k-p+q,k}^{\prime }}^\sigma (E)+\sigma \sum_{{\bf r}}F_{{\bf
k-p+q-r,r,k}^{\prime }}^\sigma (E)],  \label{f}
\end{eqnarray}
where we have performed decouplings in the spirit of ladder
approximation,
$\langle b_{-{\bf q}}^\sigma b_{{\bf q}}^{-\sigma }\rangle =N_{{\bf q}%
}^\sigma =\sigma N(\sigma \omega _{{\bf q}})$, $N(\omega )$ is
the Bose function, and $\Psi$ is defined as
\begin{equation}
\Psi _{{\bf q,k}^{\prime }}^\sigma (E)=\sum_{{\bf r}}F_{{\bf k-r,q,k}%
^{\prime }}^\sigma (E)
\end{equation}
For $U=0$ we have $G_{{\bf kk}^{\prime }}^\sigma (E)=(1-n_{{\bf k}}^\sigma
)\delta _{{\bf kk}^{\prime }}G_{{\bf k}}^\sigma (E)$, where $G_{{\bf k}%
}^\sigma (E)$ is the one-electron Green's function of the ideal crystal (cf.
Eq.(\ref{sigma})),
\begin{equation}
G_{{\bf k}}^\sigma (E)=\left[ E-t_{{\bf k}\sigma }-\Sigma _{{\bf k\sigma }%
}^{}(E)\right] ^{-1} \ , \ \ \Sigma _{{\bf k\sigma }}^{}(E)=\frac{2\overline{S}I^2Q_{%
{\bf k}}^\sigma }{1+\sigma IQ_{{\bf k}}^\sigma }  \label{g0}
\end{equation}
Note that the Eq. (\ref{g0}) gives correctly the exact Green's
function in the limit of an empty conduction band at $T=0$
\cite{Izyum,IK84,AI}.

In a general case, we have the three-particle problem (conduction
electron, core hole and magnon) which requires a careful
mathematical investigation. However, we can use the facts that the
magnon frequencies are much smaller than typical electron
energies and enrgy resolution of xas and xes methods.
Neglecting spin dynamics, the equations (\ref{g}), (\ref{f}) can
be solved exactly in a rather simple way for the case of zero
temperatures ($N_{{\bf q}}^{+}=0$, $N_{{\bf q}}^{-}=1$). Under
these conditions, $Q$ does not depend on quasimomenta, and
$\Psi_{{\bf q,k}^{\prime }}^\sigma $ does not depend on ${\bf q}$,
since the electron and magnon operator should belong to the
same perturbed site:
\begin{equation}
\Psi _{{\bf q,k}^{\prime }}^\sigma (E)=\Psi _{{\bf k}^{\prime }}^\sigma
(E)=(2S)^{1/2}\langle \langle b^\sigma c_{-\sigma }f|f^{\dagger }c_{{\bf k}%
^{\prime }\sigma }^{\dagger }\rangle \rangle _E
\end{equation}
We find in this case
\begin{eqnarray}
\Psi _{{\bf k}^{\prime }}^\sigma (E) &=&-\frac{2ISQ^\sigma (E)}{%
1+UP^{-\sigma }(E)+\sigma IQ^\sigma (E)}R_{{\bf k}^{\prime }}^\sigma (E)
\label{psi} \\
R_{{\bf k}^{\prime }}^\sigma (E) &=&\sum_{{\bf k}}G_{{\bf kk}^{\prime
}}^\sigma (E) \ , \ \ P^\sigma (E)=\sum_{{\bf k}}\frac{1-n_{{\bf k}}^\sigma }{E-t_{%
{\bf k}\sigma }}
\end{eqnarray}
After substituting Eq.(\ref{psi}) into Eq.(\ref{g}) we obtain the
following equation for the Green's function $G$
\begin{equation}
\left[ E-t_{{\bf k}\sigma }-\Sigma ^\sigma (E)\right] G_{{\bf kk}^{\prime
}}^\sigma (E)=\delta _{{\bf kk}^{\prime }}-U_{ef}^\sigma (E)\sum_{{\bf p}}G_{%
{\bf pk}^{\prime }}^\sigma (E)
\end{equation}
with the renormalized core hole potential:
\begin{equation}
U_{ef}^\sigma (E)=U\left[ 1+\frac{\Sigma ^\sigma (E)P^{-\sigma }(E)}{%
1+UP^{-\sigma }(E)+\sigma IQ^\sigma (E)}\right] .  \label{uef}
\end{equation}
Here we neglect the factor $(1-n_{{\bf k}}^\sigma )$, since the
band filling is small. Therefore one has a standard result for the
impurity scattering with renormalized energy spectrum $E_{{\bf
k}\sigma }=t_{{\bf k}\sigma }+\Sigma ^\sigma (E)$ and the
effective impurity potential $U_{ef}^\sigma (E).$ A local DOS is
given by the following expression:
\begin{equation}
N_{{\rm loc}}^\sigma (E)=-\frac 1\pi {\rm Im}G_{00}^\sigma (E)  \label{gsp}
\end{equation}
with
\begin{equation}
G_{00}^\sigma (E)=\sum_{{\bf kk}^{\prime }}G_{{\bf kk}^{\prime }}^\sigma (E)=%
\frac{R_\sigma (E)}{1+U_{ef}^\sigma (E)R_\sigma (E)}  \label{g00}
\end{equation}
where $R_\sigma (E)=\sum_{{\bf k}}G_{{\bf k}}^\sigma (E)$, and $G_{{\bf k}%
}^\sigma (E)$ is given by Eq.(\ref{g0}).

Generally speaking, theoretical investigation of the core level
spectra requires numerical calculations of realistic band
structure. We restrict ourselves to simple model calculations for
the bare semicircular DOS from the Eq.(\ref{semiel}).

The local Green function from Eq. (\ref{gsp}) describes the
absorption spectrum for $E>E_F$  and emission spectrum for
$E<E_F.$  As follows from the Eq. (\ref{gsp}), (\ref{g00}), the
experimental spectra is given by somewhat different expression
than the DOS in an initial state, and new effects can occur.

For $I>0$ the results of Eq.(\ref{uef})-(\ref{g00}) provide full
solution of the Kondo problem for an impurity in the ferromagnet,
within the parquet approximation \cite{Abr}. In the case of $I<0$,
the situation is complicated by the presence of the ``false''
Kondo divergence in the $T$-matrix \cite{Suhl}. However, this
difficulty is not important for the x-ray problem where a large
damping is always present, and experiments are performed at
sufficiently high temperatures with rather poor resolution
compared to a scale of the ``Kondo temperature''. To a leading
order in $U$ and $I$ we obtain
\begin{eqnarray}
\delta N_{{\rm loc}}^\sigma (E) &=&\frac{1-\mathop{\rm Re}(U_{ef}^{\sigma
}(E)/\Sigma ^{\sigma }(E))\left| R_\sigma ^2(E)/R_\sigma ^{\prime
}(E)\right| }{\left| 1+U_{ef}^\sigma (E)R_\sigma (E)\right| ^2}\delta
N^\sigma (E)  \nonumber \\
&&-\frac{\mathop{\rm Re}(U_{ef}^{\sigma }(E)/P^{-\sigma }(E))\left| R_\sigma
(E)\right| ^2}{\left| 1+U_{ef}^\sigma (E)R_\sigma (E)\right| ^2}\frac 1\pi
{\rm Im}P^{-\sigma }(E)  \label{nn}
\end{eqnarray}
The term in Eq.(\ref{nn}) with ${\rm Im}P^{-\sigma }(E)$ has a
smooth contribution to the spectrum. In particular, it is
non-zero in the energy gap. Note that for the emission spectra
such term is absent. The NQP contributions to the absorption (for $I>0$)
and emission spectra (for $I<0$) are proportional to $\delta N^\sigma
(E)$.

%%%%%%%%%%%%%%%%%%%% Fig.4 %%%%%%%%%%%%%%%%%%%%%%%%%%%%%%%%%%%%%%%%%%%%%%%
\begin{figure}
\centering
\includegraphics[height=5cm]{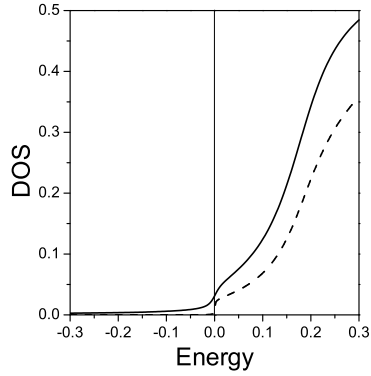}
\caption{The local density of states $N_{{\rm loc}}^{\downarrow
}(E)$ (solid line) for a half-metallic ferromagnet with
$S=1/2,I=0.3$ in the presence of the core hole potential $U=0.2.$;
smearing $E+i\delta$ is introduced with $\delta =0.01$. The dashed
line shows the DOS $N_{\downarrow }(E)$ for the ideal crystal
with spin dynamics being neglected. The value of $E_F$ calculated
from the band bottom is 0.15. The energy $E$ is referred to the
Fermi level.} \label{fig:4}
\end{figure}
\begin{figure}
\centering
\includegraphics[height=5cm]{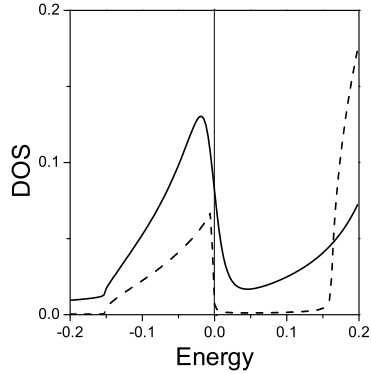}
\caption{The local density of states $N_{{\rm loc}}^{\uparrow
}(E)$ (solid line) for a half-metallic ferromagnet with
$S=1/2,I=-0.3,\protect\delta =0.025 $ in the presence of the core
hole potential $U=0.2.$ The dashed line shows the DOS
$N_{\uparrow }(E)$ for the ideal crystal. The value of $E_F$
calculated from the band bottom is 0.15.} \label{fig:5}
\end{figure}

%%%%%%%%%%%%%%%%%%%%%%%%%%%%%%%%%%%%%%%%%%%%%%%%%%%%%%%%%%%%%%%%%%%%%%%%%%
One can see from Fig.(\ref{fig:4}) that the upturn of the NQP
tail which occurs for $I>0$ becomes more sharp, although the jump
near $E_F$ weakens. For $I<0$ case, the spectral weight of NQP
contributions also increases in the presence of the core hole (see
Fig.(\ref{fig:5})). These effects have a simple physical
interpretation. Since $U_{ef}^\sigma (E)>0$ and for small band
filling $R_\sigma (E)<0$ near $E_F,$ the denominator of the
expression (\ref {nn})gives a considerable enhancement of the NQP
contributions to the spectra in comparison with those to the DOS.
However, effects of interaction $U$ turn out to be non-trivial
and do not reduce to a constant factor in the self-energy. Strong
interaction with the core hole results in a deformation of
conduction band. With increasing $U$ the spectral density
concentrated at bottom of the band. This effect is very important
for the NQP states located in this region. Therefore the spectral
weight of the NQP states increases. At very large, probably
unrealistic values of $U$, a bound state is
formed near the band bottom, and the NQP spectral weight becomes
suppressed owing to factor of $U$ in the denominator of Eq.(\ref{uef}).
%%%%%%%%%%%%%%%%%%%% Fig.5 %%%%%%%%%%%%%%%%%%%%%%%%%%%%%%%%%%%%%%%%%%%%%%%
%%%%%%%%%%%%%%%%%%%%%%%%%%%%%%%%%%%%%%%%%%%%%%%%%%%%%%%%%%%%%%%%%%%%%%%%%%

To probe a ``spin-polaron'' nature of the NQP states more explicitly, it
would be desirable to use spin-resolved spectroscopical methods such as
x-ray magnetic circular dichroism (XMCD, for a review see Ref. \cite{ebert}).
Owing to interference of electron-magnon scattering and ``exciton''
effects from interaction of electrons with the core hole, the NQP contributions
to x-ray spectra can be considerably enhanced in comparison with those to
the DOS of ideal crystal.

Now we consider the NQP effects in resonant x-ray scattering processes. It was
observed recently that the elastic peak of the x-ray scattering
in CrO$_2$ is more pronounced than in usual Cr compounds \cite{kurm}.
The authors of this work have put forward some
qualitative arguments that the NQP states may give larger contributions to
resonant x-ray scattering than usual itinerant electron states. Here we
shall treat this problem quantitatively and estimate explicitly the
corresponding enhancement factor. The intensity of resonant x-ray emission
induced by the photon with the energy $\omega $ and polarization \thinspace $%
q$ is given by the Kramers-Heisenberg formula \cite{sakurai,yab,gelm}
\begin{equation}
I_{q^{\prime }q}(\omega ^{\prime },\omega )\propto \sum_n\left| \sum_l\frac{%
\langle n|C_{q^{\prime }}|l\rangle \langle l|C_q|0\rangle }{E_0+\omega
^{\prime }-E_l-i\Gamma _l}\right| ^2\delta (E_n+\omega ^{\prime }-E_0-\omega
),  \label{xray}
\end{equation}
here $q^{\prime },\omega ^{\prime }$ are the polarization and energy of the
emitted photon, $|n\rangle$, $|0\rangle $ and $|l\rangle $ are the final,
initial and intermediate states of the scattering system with the energies $E_i$, respectively,
and $C_q$ is the operator of a dipole moment
for the transition, which is proportional to $(fc+c^{\dagger }f^{\dagger })$.
For simplicity we assume hereafter that $\Gamma _l$ does not depend on
the intermediate state: $\Gamma _l=\Gamma $, and take into account only the
main x-ray scattering channel where the hole is filled by conduction electron.
Assuming that the electron-photon interaction that induces the
transition is contact, the expression for threshold scattering intensity
has following form \cite{sokolov}
\begin{eqnarray}
I_{\omega ^{\prime }} &\propto &\sum_{\sigma \sigma ^{\prime }}\int_0^\infty
dt_1\int_0^\infty dt_2\exp \left[ -i(\omega ^{\prime }-\varepsilon
_0)(t_1-t_2)-\Gamma (t_1+t_2)\right]  \nonumber \\
&&\langle 0|c_\sigma \exp (i{\cal H}_ft_1)c_{\sigma ^{\prime }}^{\dagger
}\exp [i{\cal H}_i(t_2-t_1)]c_{\sigma ^{\prime }}\exp (-i{\cal H}%
_ft_2)c_\sigma ^{\dagger }|0\rangle , \label{ffff}
\end{eqnarray}
where $H_f$ and $H_i$ are conduction-electron Hamiltonians with and without
core hole, respectively. The complicated correlation function in Eq.(\ref{ffff})
can be decoupled in the ladder approximation which is exact for the empty
conduction band. Then one can obtain \cite{sokolov}
\begin{equation}
I_{\omega ^{\prime }}\propto \left| \sum_\sigma G_{00}^\sigma (z)\right| ^2 ,
\label{zz}
\end{equation}
where $z=\omega ^{\prime }-E_0+i\Gamma .$ Owing to a jump in the
DOS at the Fermi level, the NQP part of the Green's function
contains a large logarithm $\ln (W/z)$ at small $z$, $W$ being a
bandwidth. It means that the corresponding contribution to the
elastic x-ray scattering intensity ($\omega ^{\prime }=E_0$) is
enhanced by a factor of $\ln ^2(W/\Gamma )$, which makes a
quantitative estimation for the qualitative effect discussed in
Ref. \cite{kurm}. Of course, the smearing of the jump in the
density of NQP states by spin dynamics is irrelevant provided
that $\Gamma > \overline{\omega }$, where $\overline{\omega}$
 is a characteristic magnon frequency.

\section{Transport properties}

Transport properties of the HMF are a subject of numerous experimental
investigations (see, e.g., recent works on CrO$_2$ \cite{CrO2},  NiMnSb
\cite{NiMnSb}, and the reviews \cite{IK,ziese,nagaev}). At the same time,
a theoretical interpretation of these results is still problematic. Concerning
electronic scattering mechanisms, the most important difference between the HMF
and ``standard'' itinerant electron ferromagnets like iron or nickel is the
absence of one-magnon scattering processes in the former case \cite{IK}.
Two-magnon scattering processes have been considered many years ago for both
the broad-band case (a weak $s$-$d$ exchange interaction) \cite{Ros} and
narrow-band case (a ``double exchange model'') \cite{ohata}. Obtained
temperature dependence of resistivity have the form $T^{7/2}$ and $%
T^{9/2}$, respectively. At low enough temperatures the first result fails
and should be replaced by $T^{9/2}$ as well \cite{lutovinov}; the reason is
a compensation of transverse and longitudinal contributions in the
long-wavelength limit, which is a consequence of the rotational symmetry of
the $s$-$d$ exchange Hamiltonian \cite{nagaev1,AKI}. Recently a general
interpolation theory has been formulated \cite{ourtransport}. Here we
discuss main results of this work with a special emphasize to the NQP effects.

In the spin-wave region the Hamiltonian (\ref{H}) can be rewritten in the
form
\begin{eqnarray}
{\cal H} &=&{\cal H}_0-I(2S)^{1/2}\sum_{{\bf kq}}(c_{{\bf k}\uparrow
}^{\dagger }c_{{\bf k}+{\bf q}\downarrow }b_{{\bf q}}^{\dagger }+h.c.)
\nonumber \\
&&\ \ \ \ \ +I\sum_{{\bf kqp}\sigma }\sigma c_{{\bf k}\sigma }^{\dagger }c_{%
{\bf k+q-p}\sigma }b_{{\bf q}}^{\dagger }b_{{\bf p}}
\end{eqnarray}
Here the zero-order Hamiltonian includes non-interacting electrons and
magnons:
\begin{equation}
{\cal H}_0=\sum_{{\bf k}\sigma }t_{{\bf k}\sigma }c_{{\bf k}\sigma
}^{\dagger }c_{{\bf k}\sigma }+\sum_{{\bf q}}\omega _{{\bf q}}b_{{\bf q}%
}^{\dagger }b_{{\bf q}},
\end{equation}
with the spin splitting $\Delta =2IS$ being included in ${\cal H}_0$. In the
half-metallic case spin-flip processes do not appear in the second order
in $I$, since the states with only one spin projection presented at the
Fermi level. At the same time, we have to consider the renormalization of
the longitudinal processes in higher orders in $I$ (formally, we need to
include all terms up to the second order in a quasiclassical small
parameter $1/S$). To this end we eliminate from the Hamiltonian the terms
which are linear in the magnon operators by using the canonical
transformation \cite{nagaev1}. Then, the effective Hamiltonian has a following form
\begin{equation}
\widetilde{{\cal H}}={\cal H}_0+\frac 12\sum_{{\bf kqp}\sigma }({\cal A}_{%
{\bf kq}}^\sigma +{\cal A}_{{\bf k+q-p,q}}^\sigma )c_{{\bf k}\sigma
}^{\dagger }c_{{\bf k+q-p}\sigma }b_{{\bf q}}^{\dagger }b_{{\bf p}} ,
\label{hef}
\end{equation}
where
\begin{equation}
{\cal A}_{{\bf kq}}^{\sigma }=\sigma I\frac{t_{{\bf k+q}}-t_{{\bf k}}}{t_{%
{\bf k+q}}-t_{{\bf k}}+\sigma \Delta }  \label{amp}
\end{equation}
is the $s$-$d$ scattering amplitude, which vanishes at $q\rightarrow 0$ and
thereby takes properly into account the rotational symmetry of
electron-magnon interaction. More general interpolation expression for the
effective amplitude which does not assume the smallness of $|I|$ or $1/S$
was obtained in Ref. \cite{AKI} within a variational approach, but it does not
differ qualitatively from simple expression (\ref{amp}). In the case of real
itinerant magnets including the HMF, a ${\bf k}$-dependence of $s$-$d$ exchange
parameter should be taken into account, similarly to the
temperature dependence of spin polarization. However, here we restrict
ourselves only to the rigid spin splitting model appropriate for degenerate ferromagnetic
semiconductors. One can expect from phenomenological symmetry considerations
that the temperature dependences of transport properties are rather universal.

The most general scheme for calculating the transport
relaxation time is the Kubo formalism for the conductivity $\sigma_{xx}$ \cite{kubo}
\begin{equation}
\sigma _{xx}=\beta \int_0^\beta d\lambda \int_0^\infty dt\exp (-\varepsilon
t)\langle j_x(t+i\lambda )j_x\rangle  \label{kru}
\end{equation}
where $\beta =1/T,$ $\varepsilon \rightarrow 0$, ${\bf j}=-e\sum_{{\bf k}%
\sigma }{\bf v}_{{\bf k}\sigma }c_{{\bf k}\sigma }^{\dagger }c_{{\bf k}%
\sigma }$ is the current operator, ${\bf v}_{{\bf k}\sigma }=\partial t_{%
{\bf k}\sigma }/\partial {\bf k\ }$ is the electron velocity. Rewriting
the total Hamiltonian in the form ${\cal H}={\cal H}_0+{\cal H}_1$, the
correlator in (\ref{kru}) may be expanded in the perturbation ${\cal H}_1$
\cite{Nak}. In the second order we obtain for the electrical resistivity
the following expression
\begin{equation}
\rho _{xx}=\sigma _{xx}^{-1}=\frac T{\langle j_x^2\rangle ^2}\int_0^\infty
dt\langle [j_x,{\cal H}_1(t)][{\cal H}_1,,j_x]\rangle ,
\end{equation}
where ${\cal H}_1(t)$ is calculated with the Hamiltonian ${\cal H}_0$. In
the HMF situation the band states with one spin projection only, $\sigma
=\alpha ={\rm sign}I,$ are present at the Fermi level. Below we consider the
case $I>0,$ $\sigma =+$ and omit the spin indices in the electron spectrum.
Then one can find an expression for the transport relaxation time $\tau $ defined as $\sigma
_{xx}=e^2\langle (v^x)^2\rangle \tau $
\begin{eqnarray}
\frac 1\tau &=&\frac \pi {4T}\sum_{{\bf kk}^{\prime }{\bf q}}(v_{{\bf k}%
}^x-v_{{\bf k}^{\prime }}^x)^2({\cal A}_{{\bf kq}}^{\uparrow }+{\cal A}_{%
{\bf k}^{\prime }{\bf ,q-k}^{\prime }{\bf +k}}^{\uparrow })^2N_{{\bf q}%
}(1+N_{{\bf q-k}^{\prime }+{\bf k}})
 n_{{\bf k}}(1-n_{{\bf k}^{\prime }})   \nonumber \\
&\times& \delta (t_{{\bf k}^{\prime }}-t_{{\bf k}}-\omega _{{\bf q}%
}+\omega _{{\bf q-k}^{\prime }+{\bf k}})\left/ \sum_{{\bf k}}(v_{{\bf k}%
}^x)^2\delta (t_{{\bf k}})\right.  \label{tau}
\end{eqnarray}
Averaging over angles of the vector ${\bf k}$ leads us to the final result
$1/\tau \propto I^2\Lambda $ with
\begin{equation}
\Lambda =\sum_{{\bf pq}}f_{{\bf pq}}\frac{\beta (\omega _{{\bf p}}-\omega _{%
{\bf q}})|{\bf p-q}|}{\exp \beta \omega _{{\bf p}}-\exp \beta \omega _{{\bf q%
}}}(1+N_{{\bf q}})(1+N_{{\bf p}}) , \label{lam}
\end{equation}
where $f_{{\bf pq}}=1$ for $p,q\gg q_0$ and
\begin{equation}
f_{{\bf pq}}=\frac{[{\bf p\times q]}^2}{({\bf p-q)}^2q_0^2}\,\,\,(p,q\ll
q_0).
\end{equation}
The wavevector $q_0$ determines the boundary of a region where ${\bf q}$%
-dependence of the amplitude become important, so that $t({\bf k+q})-t({\bf k%
})\simeq \Delta $ at $q\simeq q_0$ and the simple perturbation theory fails.
In elementary one-band model of the HMF where $E_F<\Delta $ one has $q_0\sim
\sqrt{\Delta /W}$ (where $W$ is the conduction bandwidth, and the lattice constant is put
to unity) \cite{nagaev1}. Generally speaking, $q_0$ may be sufficiently
small provided that the energy gap is much smaller than $W$, which is the
case for real HMF systems.

The quantity $q_0$ determines a characteristic temperature and energy scale $%
T^{*}=Dq_0^2\propto D({\Delta /}W)$, where $D\propto T_C/S$ is the spin-wave
stiffness defined by $\omega _{{\bf q\rightarrow }0}=Dq^2$, and $T_C$ is the
Curie temperature. It is important that similar crossover temperatures
appears in the temperature dependence of the spin polarization (see, e.g.,
Eqs.(\ref{phi4}),(\ref{phi11})). This means that temperature dependences of
both spin polarization and transport properties can be changed at low enough
temperatures within the spin-wave temperature region.

One has to bear in mind that each power of $p$ or $q$ yields the $T^{1/2}$ factor
for temperature dependence of resistivity. At very low
temperatures $T<T^{*}$ small quasimomenta $p,q<q_0$ gives a main contribution
to the integrals. Then the temperature dependence of  resistivity is equal to $\rho (T)\propto
(T/T_C)^{9/2}.$ Such a dependence was obtained in the large-$|I|$ case where
the scale $T^{*}$ is absent \cite{ohata}, and within a diagram approach in the
broad-band case \cite{lutovinov}. At the same time, for $T>T^{*}$ the
function $f_{{\bf pq}}$ in Eq. (\ref{lam}) can be replaced by unity, leading to
$\rho (T)\propto (T/T_C)^{7/2},$ in agreement with the old results \cite
{Ros}.

According to calculations presented here, the NQP
states do {\it not} contribute to the temperature dependence of the
resistivity for pure HMF. An opposite conclusion was made by Furukawa \cite
{furukawa} and related to an anomalous $T^3$ dependence in the resistivity.
However, this calculation was not based on a consistent use of the Kubo formula and,
in our opinion, can be hardly justified.

On the contrary, {\it impurity} contributions to transport properties in the
presence of potential scattering are determined mainly by the NQP states (it
has been shown first in Ref. \cite{IKT}, see also Ref. \cite{IK}). To a second order in
the impurity potential $V$ we derive for the electron Green's function
\begin{eqnarray}
G_{{\bf kk}^{\prime }\sigma }(E) =\delta _{{\bf kk}^{\prime }}G_{{\bf k}%
\sigma }^{(0)}(E)+G_{{\bf k}\sigma }^{(0)}(E)VG_{{\bf k}^{\prime }\sigma
}^{(0)}(E)
[1+V\sum_{{\bf p}}G_{{\bf p}\sigma }^{(0)}(E)] ,
\end{eqnarray}
where $G_{{\bf k}\sigma }^{(0)}(E)$ is the exact Green's function for the
ideal crystal (see Eq.(\ref{dys})). Neglecting vertex corrections and averaging
over impurities, we obtain for the transport relaxation time in the following form
\begin{equation}
\delta \tau _{imp}^{-1}(E)=-2V^2{\rm Im}\sum_{{\bf p}}G_{{\bf p}\sigma
}^{(0)}(E)
\end{equation}
Thus the relaxation time is determined by the energy
dependence of the density of states $N(E)$ for the interacting system near
the Fermi level. The most nontrivial dependence comes from the
nonquasiparticle states with the spin projection $\alpha ={\rm sign}I$,
which are present near the $E_F$. Close to the Fermi level the NQP
contribution follows the power law (\ref{alpha1}). Therefore, the
impurity contribution to the resistivity  is equal to
\begin{eqnarray}
\frac{\delta \rho _{imp}(T)}{\rho ^2} =-\delta \sigma _{imp}(T)
\propto -V^2\int dE\left( -\frac{\partial f(E)}{\partial E}\right)
\delta N_{incoh}(E)\propto T^{3/2}  \label{rimp}
\end{eqnarray}
The contribution of the order of $T^\alpha $ with $\alpha \simeq 1.65$
(which is not too far from 3/2) has been observed recently in the
temperature dependence of the resistivity for NiMnSb \cite{NiMnSb}.

To calculate the magnetoresistivity we take into account a gap in the magnon
spectrum induced by magnetic field, $\omega _{{\bf q\rightarrow }0}=Dq^2+\omega _0$.
For large external magnetic field $H$, in comparison with the anisotropy gap,
$\omega _0$ is proportional to $H$ . At $T<T^{*}$ the resistivity is linear
in magnetic field:
\begin{equation}
\rho (T,H)-\rho (T,0)\propto -\omega _0T^{7/2}/T_C^{9/2}
\end{equation}
The situation at $T>T^{*}$ is more interesting since the quantity $\partial
\Lambda /\partial \omega _0$ contains a logarithmic divergence with the cutoff
at $\omega _0$ or $T^{*}$. We have at $T>\omega _0$ , $T^{*}$:
\begin{equation}
\delta \rho (T,H)\propto -\frac{T^3\omega _0}{[\max (\omega _0,T^{*})]^{1/2}}
\end{equation}
Of course, at $T<\omega _0$ the resistivity is exponentially small. A
negative $H$-linear magnetoresistance was observed recently in CrO$_2$ \cite
{CrO2}. The incoherent contribution to magnetoresistivity is given by
\begin{equation}
\delta \rho _{imp}(T,H)\propto \omega _0\partial \delta N_{incoh}(\sigma
T)/\partial T\propto \omega _0\sqrt{T}.
\end{equation}

Another useful tool to detect the NQP states is provided by tunneling phenomena
\cite{AI1}, in particular by the Andreev reflection spectroscopy for the
HMF-superconductor tunnel junction \cite{falko}. A most direct way is
the measurement of a tunnel current between two pieces of the HMF with
the opposite magnetization directions. To this end we consider a standard
tunneling Hamiltonian (see, e.g., Ref. \cite{Mah}, Sect. 9.3):
\begin{equation}
{\cal H}={\cal H}_L+{\cal H}_R+\sum_{{\bf kp}}(T_{{\bf kp}}c_{{\bf k}%
\uparrow }^{\dagger }c_{{\bf p}\downarrow }+h.c.),
\end{equation}
where ${\cal H}_{L,R}$ are the Hamiltonians of the left (right) half-spaces,
respectively, ${\bf k}$ and ${\bf p}$ are the corresponding quasimomenta,
and spin projections are defined with respect to the magnetization direction
of a given half-space (the spin is supposed to be conserving in the ``global''
coordinate system). Carrying out standard calculations of the tunneling
current ${\cal I}$ in the second order in $T_{{\bf kp}}$ we obtain (cf. Ref. \cite
{Mah})
\begin{eqnarray}
{\cal I} \propto \sum_{{\bf kqp}}|T_{{\bf kp}}|^2[1+N_{{\bf q}}-f(t_{{\bf %
p-q}})]
[f(t_{{\bf k}})-f(t_{{\bf k}}+eV)]\delta (eV+t_{{\bf k}}-t_{{\bf p-q%
}}+\omega _{{\bf q}}) \nonumber
\end{eqnarray}
Here $V$ is the bias voltage. For $T=0$ one has
$
d{\cal I}/dV\propto \delta N_{incoh}(eV).
$

\section{Conclusions}

To conclude, we have considered the special properties of half-metallic
ferromagnets which are connected with their unusual electronic structure.
Further experimental investigations would be of a great
importance, especially keeping in mind possible role of the HMF for different applications
\cite{IK,pickett,prinz}.

Several experiments could be performed in order to clarify the impact of the
nonquasiparticle states on spintronics. Direct ways of observing the NQP
states would imply the technique of Bremsstrahlung Isohromat Spectroscopy
(BIS) \cite{BIS} or the spin-polarized scanning tunneling microscopy (SP-STM) \cite{STM},
since for the most frequent case of minority-spin gap where the
NQP states lie above $E_{F}$. In contrast with the photoelectron
spectroscopy of the occupied states (PES) which has to show a complete spin
polarization in the HMF with minority-spin gap, the BIS spectra should demonstrate
an essential depolarization of the states above the $E_{F}$. For    the
majority-spin-gap HMF, vice versa, the partial depolarization should be seen
in the PES. The $I-V$ characteristics of half-metallic tunnel junctions for the case
of antiparallel spins are completely determined by the NQP states \cite{ourtransport,falko1}.
The spin-polarized STM should be able to probe these
states by the differential tunneling conductivity $dI/dV$ \cite{Mah,tun}.
In particular, the SP-STM with
positive bias voltage can detect the opposite-spin states just above the
Fermi level for surface of the HMF such as CrO$_{2}$. The Andreev reflection
spectroscopy for tunnel junction superconductor-HMF \cite{falko} can
also be used in searching for experimental evidence of the NQP effects. These
experimental measurements will be of crucial importance for the theory of
spintronics in any tunneling devices with the HMF. Since ferromagnetic
semiconductors can be considered as a special case of the HMF, an account of
these states can be helpful for the proper description of spin diodes and
transistors \cite{vignale}.

The research described was supported in part by Grant No.02-02-16443 from
Russian Basic Research Foundation and by Russian Science Support Foundation.

\end{document}